# Multiple faults diagnosis using causal graph

*Imtiez Fliss and Moncef Tagina*

UNIVERSITY OF MANOUBA, NATIONAL SCHOOL OF COMPUTER SCIENCES, SOIE LABORATORY

2010 MANOUBA, TUNISIA

*{Imtiez.Fliss, Moncef.Tagina} @ ensi.rnu.tn*

**Abstract**

This work proposes to put up a tool for diagnosing multi faults based on model using techniques of detection and localization inspired from the community of artificial intelligence and that of automatic. The diagnostic procedure to be integrated into the supervisory system must therefore be provided with explanatory features. Techniques based on causal reasoning are a pertinent approach for this purpose. Bond graph modeling is used to describe the cause effect relationship between process variables. Experimental results are presented and discussed in order to compare performance of causal graph technique and classic methods inspired from artificial intelligence (DX) and control theory (FDI).

## 1. INTRODUCTION

With the continuous expansion of industrial applications, complex systems diagnosis has become an extremely discussed topic nowadays.

Many methods have been proposed by different communities (artificial intelligence, control theory, statistics...). Several issues are increasingly asked, particularly regarding diagnosing faults. The general ones concern diagnosing single fault. Other questions tackle more complex points and present the problem of detection and localization of multiple faults (Multiple faults means that multiple breakdowns have effects that overlap in time).

This problem occurs in many industrial systems. However, the number of research in this field is reduced due to the complexity of the task and combinatorial explosion problem.

In our work, we propose to build a multi faults diagnosing tool based on model approach. We use techniques of detection and localization inspired from the community of artificial intelligence and that of automatic. The diagnostic procedure to be integrated into the supervisory system must therefore be provided with explanatory features. Techniques based on causal reasoning are a pertinent approach for this purpose.

In this paper, we intend to present in the second section causal reasoning which is the base of our multiple faults diagnosing tool. Then, we will show and discuss the experimental results provided by our diagnosing tool in order to compare performance of causal graph technique and classic methods inspired from artificial intelligence and control theory.

## 2. CAUSAL REASONING: THE STATE OF THE ART

The diagnosis of complex systems has been an area of dynamic research for many years.

Two research communities have been particularly involved in studying fault diagnosis: the artificial intelligence community, known as the diagnostic (DX) community, and the control theory community, known as the fault detection and isolation (FDI) community.

The DX community has been concerned with the modeling of the diagnostic reasoning itself: the foundations of logical reasoning have always been considered as major research points. In the consistency based approach [18], the description of the behavior of the system is component-oriented and rests on first-order logic. The {SD (system description), COMP (components)} pair constitutes the model. The system description takes the form of logical operations [4].

Diagnosis in this framework is logically sound but a major drawback is the issue of combinatorial explosion for systems involving many components [3], as in the case of industrial processes, with whose diagnosis this paper is concerned. Reference [3] also, states that one of the main limitations in logical model based diagnosis is its computational complexity, and proposes a specific knowledge compilation approach to focus reasoning on abductive diagnosis.

The FDI community is especially concerned with industrial process modeling and control. Reasoning is quantitative. [10]

The model is numerical and as precise as required by the diagnostic objective. Generally, the model represents the normal behavior of the system, in the absence of any fault and characterizes deterministic phenomena, taken into consideration using basic "laws" of physics, biology, etc. But it can also include detailed knowledge about how faults or unknown disturbances affect the variables of the system.

With the FDI approach, computations result in numerical quantities, the residuals, whose properties enable diagnosis with very accurate quantitative information.

Nevertheless, several logical assumptions are implicit in the FDI formulation, whereas they are clearly formulated within the DX consistency-based diagnostic framework.

Reasoning and computing may thus be considered in opposition. A combined method presented in [9] brings them together. It relies on both a qualitative causal representation of the process and quantitative local models. It has been inspired by artificial intelligence for the causal modeling of physical systems and for studying logical soundness. It is causal reasoning in which we are interested in this paper.

A causal structure is a qualitative description of the effect or influence that system entities (variables, faults, etc.) have on other entities. It may be represented by a directed graph (digraph). A causal graph, which represents a process at a high level of abstraction, is appropriate for supervising the process. When the graph nodes represent the system variables, the directed arcs symbolize the normal relations among them and these relations are deterministic, the graph is frequently referred to as an influence graph. [9]

Diagnosis based on influence graph consists in seeking for the variable source which deviation is sufficient to explain all deviations detected on other variables. [20]

The algorithm is a backward/forward procedure starting from an inconsistent variable. The backward search bounds the fault space by eliminating the normal measurements causally upstream. Then each possible primary deviation generates a hypothesis, which is forward tested using the states of the variables and the functions of the arcs. [9]

Localization phase consists in looking for the system component that doesn't correctly work using the system structure knowledge, its potential weaknesses and available observations.

The result of diagnosis may be an arc pointing on a source variable (component fault) or a non measurable disturbance that directly affects that variable.

A major advantage of causal approach is that, in general, faulty behavior knowledge is not necessary for localization.

Two principle causal structures are proposed:

a) Digraph that represents calculability issued from mathematic relations (differential equation ...). It may be found through the causal-ordering mechanism [17] or through the graph bipartite theory.

b) Digraph that represents functional knowledge of process; nodes are linked to significant considered variable and arcs are linked to physical phenomena [7].

Thus, the first type makes a link between causality and system describing equations (global analysis) and the second one makes a link between causality and system structure (local analysis).

We have also bond graph -a many diagnosis approaches' base- such as temporal causal graph [16]

The first step for causal graph diagnosing is the construction of the causal graph. This is a complex process that needs a structural and functional knowledge.

Expert knowledge is also considered to define supervision needs. Reference [12] lists some tips to build causal graph within a complex system context:

i. Physical system identification.

i. Dividing physical system to subsystems.

ii. Defining and affecting a configuration to every subsystem.

iii. Identification of some physical relations.

iv. Connecting relations to physical components.

v. Causality determination.

vi. Reducing (eliminating non measurable variables).

vii. Approximation (eliminating negligible relations).

viii. Quantification (identification of transfer function parameters).

Many researches are based on causal graph. The principle approaches are numerated in [20]. In this paper, we focus on some approaches.

### 2.1 Causal Engine (Ca-En)'s Approach:

Ca-En is a qualitative simulator developed within the European project ESPRIT TIGER. It is a model based diagnosis system for complex dynamic process, integrated in the supervision system of gas turbine TIGER.

The CA-EN formalism is based on a two level representation scheme for describing the relationships between the process variables: a local constraint level and a global constraint level. [21]

The local constraint level is represented by a directed graph in which the paths presume the perturbation flow causality. The influences supported by the graph's edges allow for representing causal dependency type knowledge.

The global constraint level is composed of functional numeric constraints associated with interval domains, such as constraints arising from physical laws. So, a global constraint is any mathematical equation, which might be nonlinear as well, in which each unknown is assumed to take on interval values. [21]

The "Causalito" program automatically performs the difficult part of translating analytic knowledge into causal relations. "Causalito" uses causal-ordering concepts [17] for automatically generating the CA-EN causal graph as well as some of the influence attributes from a set of equations.

Imprecise knowledge is considered through the definition of intervals on relation parameters (associated to influences). This allows prediction envelop generation and updates every Sampling period.

## 2.2 Evsukoff 's Approach

Supervision of complex process is also dealt by Alexandre Evsukoff [8]. He Suggests a causal approach similar to the one proposed by CA-EN. The difference lies in detection mechanism and localization process.

Detection is based on fuzzy inference on the residual attributes where the residual (r) is the difference between process measurements y (t) and referential values ŷ (t) issued from the model. A study on the robustness and sensitivity of detection is also performed.

The uncertain reasoning based on intervals and envelopes in Ca-EN is then replaced by a fuzzy reasoning on residuals. That allows a more refined reasoning on the differences that occur.

Localization is based on causal reasoning: each variable is physically linked to other variables that cause and explain its behavior. It is based on a multi-model that defines for each variable global, causal (local), and propagated residual.

The localization procedure is applied generally to every variable of the process. The mechanism of inference is made on each residual. A decision localization process verify (for each variable in alarm) if disturbance is detected locally or not.

## 3. BOND GRAPH MODELING

To ensure the best precision and reliability in detection and isolation, we have to choose carefully the model of the physical system. In fact, the quality of diagnostic system depends on the quality of the model.

A model is a simple or abstract representation (diagram, graphic representation, mathematic equations, etc) of a physical system. Dynamic models of physical systems may be represented in different ways: logical statements [11], mathematic equations [6], [13], bond graph [15], bloc diagram and bond graph [22], digraphs …

The preference of the adequate representation of the physical system depends on the purpose of the search.
In our case, we focus on bond graph modeling [5].
In fact, Bond graph language allows to deal with the enormous amount of equations describing the process behavior and to display explicitly the power exchange between the process components starting from the instrumentation architecture. It is a unified language for all engineering science domains that considers energy and information channels. Indeed, that is very useful since multidisciplinary systems constitute the majority of industrial products that exist nowadays.

The causality, which establishes the cause and effect relationship between the power variables, is an important characteristic used in bond graph models to derive the constitutive equations of the process behavior in a systematic and an algorithmic way. The verification of the causality assignment avoids design and numerical simulation problems.

## 4. SYSTEM AND MODELS PRESENTATION

As we presented at the beginning of this paper, our work consists in establishing a diagnosis system for multiple faults based on causal reasoning. We use for this purpose the method of influence graph for isolating faults described in [9]. In fact, causal structure of influence graphs provides a tool to know and understand how normal or abnormal variations propagate in the physical process from one variable to another [10].

This allows us to know the state of components even in case of multiple faults (our study case).
Then, the results are compared with those given by FDI and Logical method with fault models (DX) in case of multiple faults.

To test the performance of the proposed methods, we have chosen a benchmark in diagnosis domain: three tanks hydraulic system [14] [19] [1] …
Fig. 1 illustrates the notation used in this section. The process consists of three cylindrical tanks. Tanks communicate through feeding valves. The process has two inputs: Msf1 and Msf2. We put five sensors: effort sensors De1, De2 and De3 to measure pressure of C1, C2 and C3 and flow sensors Df1 and Df2 measuring flow level of the valves 1 and 2. Its global purpose is to keep a steady fluid level in the tanks.

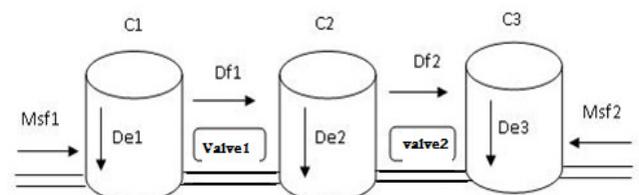

Fig 1. Three-tank system

Then, we used a procedure described in [2] and [5] that let us get the bond graph model of the process shown in Fig2

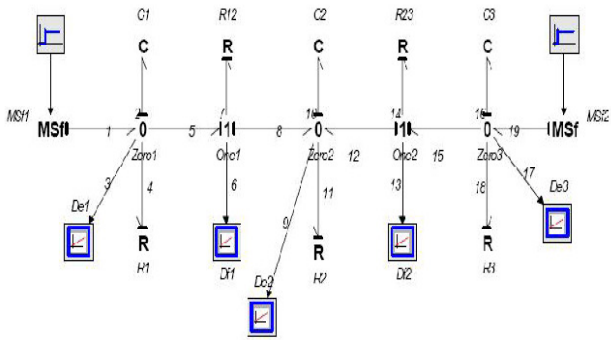

Fig 2 . Bond Graph model of three tanks system.

Thanks to structural, behavioral and causal properties of Bond Graph, the causal graph of the three tanks process can be generated as given in Fig3.

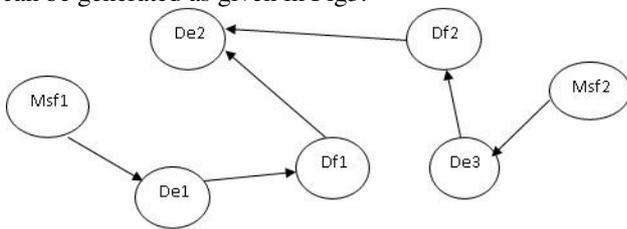

Fig 3. Influence Graph of three - tank system

In this case, the graph nodes represent the system variables; the directed arcs symbolize the normal relations among them (for instance Msf1->De1 means that modifications of Msf1 will necessary cause changes of De1).

## 5. IMPLEMENTATION AND DISCUSSION

The multi faults diagnosis tool we put up is original. It allows us to compare the results of three techniques inspired from the community of artificial intelligence and control theory: FDI, Logical method with fault models and influence graph. The experimental results were obtained during the simulation of the three - tank system.

After generating the influence graph, the propagation paths in the graph are analyzed to determine whether this fault hypothesis is sufficient to account for secondary faults, resulting from its propagation in the process over time. The algorithm is a backward/forward procedure starting from an inconsistent variable (source variable). The backward search bounds the fault space by eliminating the normal measurements causally upstream. Then each possible primary deviation generates a hypothesis, which is forward tested using the states of the variables and the functions of the arcs.

The results of this approach and those of FDI and Logical method with fault models (DX) tested on the three- tank process are presented in the following table:

| Injected faults | FDI | Logical method with fault models (DX) | Influence Graph |
|---|---|---|---|
| {Msf1} | {Msf1} | {Msf1} | {Msf1} |
| {Msf2} | {Msf2} | {Msf2} | {Msf2} |
| {De1} | {De 1} | {De 1} | {De 1} |
| {De2} | {De2} | {De2} | {De2} |
| {De3} | {De1, f2} {Df2, f1} {De3, Df1} | {De3} | {De3} |
| {Df1} | {Df1} | {Df1} | {Df1} |
| {Df2} | {De1, f2} {Df2, Df1} {De3, Df1} | {Df2} | {Df2} |
| {Msf1, Df2} | {De1, f2} {Df2, Df1} {De3, Df1} | {Df2} | {Msf1,Df2} |
| {De1, Df2} | {De1, Df2} {Df2, f1} {De3, Df1} | {De1, Df2} | {De1, Df2} |
| {De3, Df2} | { } | {De3, Df2} | {De3, Df2} |
| {De1, De3} | {De1, Df2} {Df2, Df1} {De3, Df1} | {De1, De3} | {De1, De3} |
| {Df1, Df2} | {De1, Df2} {Df2, f1} {De3, Df1} | {Df1, Df2} | {Df1, Df2} |
| {Msf1, De1, Df2} | {De1, Df2} {Df2, Df1} {De3, Df1} | {De1, Df2} | {Msf1, De1, Df2} |
| {De1, De3, Df2} | {Df1, Msf2} | {De1, De3, Df2} | {Df1, Df2, De3} |
| {De1, Df1, Df2} | {De1, Df2} {Df2, Df1} {De3, Df1} | {De1, Df1, Df2} | {De1, Df1, Df2} |
| {Msf1, Msf2, Df1} | { } | {Msf2, Df1} | {Msf1, Msf2, Df1} |
| {Msf1, Msf2, Df1, Df2} | {De1, Df2} {Df2, Df1} {De3, Df1} | {Df1, Df2} | {Msf1, Msf2, Df1, Df2} |
| {Msf1, Msf2, Df1, Df2, De2} | {De1, Df2} {Df2, Df1} {De3, Df1} | {De2, Df1, Df2} | {Msf1, Msf2, De2, Df1, Df2} |
| {Msf1, Msf2, De1, De2, Df1, Df2} | {De1, Df2} {Df2, Df1} {De3, Df1} | {De1, De2, Df1, Df2} | {Msf1, Msf2, De1, De2, Df1, Df2} |

Tab 1. Experimental results

These results are better presented in the chart below.

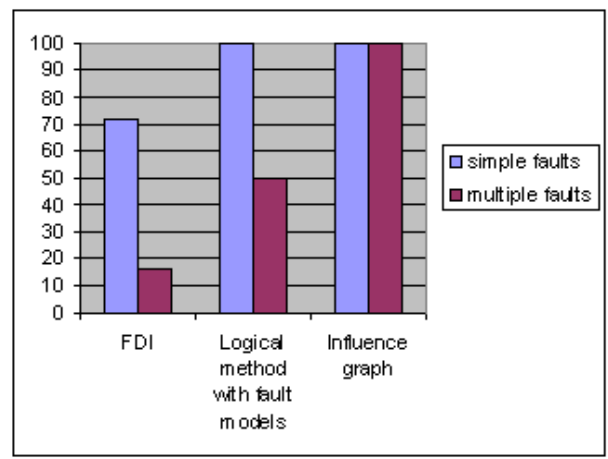

Fig 4. Experimental results

Based on the experimental results, we notice that the three techniques give good results in the majority of cases of a simple fault (the three techniques localized more than 71% of simple injected faults).

However, in multiple faults instances (double and more), the three methods give different even don't give results. FDI localized only 16.6% of double and more faults.

This can be explained by the fact that the generation and use of theoretical fault signatures reduce the diagnostic reasoning to a simple pattern-matching activity (this matches with considerations of [9]).

Logical method with fault models gives results better than FDI: 50% of multiple injected faults (because the generated diagnosis is revised using fault model technique) but not than Influence Graph (as the diagnosis initially generated is minimal since it is a result of HS tree).

On the other hand, Influence Graph method gives very interesting results localizing perfectly faults. The registration of all information concerning variables may explicate these results. Then, only variables that are really faulty are announced defective.

To conclude, associating causal approaches would be an interesting solution for dynamic complex systems since they limit verification space of diagnosis system to relations sufficient to isolate faults and give remarkable end results in complex case: multiple faults without a need for supplementary processing.

## 6. CONCLUSION

This paper has presented multiple fault isolation method based on causal reasoning. Bond Graph modeling was used to describe the relationship cause effect existing between process variables. A comparison between results giving by different approaches: FDI, logical method with fault models and influence graph was done for explanation and implementation purposes.

Experiments have shown that the causal reasoning through the example of influence graph can localize multiple faults in the three-tank process successfully. It is expected that the achieved results can also be extended to localizing multiple faults in genuine systems.
In fact, we intend in future works to highlight the potential of using such a method in real application.